\documentclass[twocolumn,showpacs,preprintnumbers,amssymb]{revtex4}
\usepackage{graphicx, epsfig, amssymb} 
\usepackage{amsmath, amsfonts}
\usepackage{times}
\usepackage{bm} 
\usepackage[usenames]{color}

\usepackage[tracking=true]{microtype}
\SetTracking{}{500}
\SetTracking{encoding={*}, shape=sc}{40}
\begin{document}  
\title{Overcharging higher-dimensional black holes with point particles}

\author{Karl Simon Revelar,}
\email{karevelar@upd.edu.ph}
\author{Ian Vega}
\email{ivega@nip.upd.edu.ph}
\affiliation{National Institute of Physics, University of the Philippines, Diliman, Quezon City 1101, Philippines}

\begin{abstract}
We investigate the possibility of overcharging spherically-symmetric black
holes in spacetime dimensions $D > 4$ by the capture of a charged particle. We generalize Wald's classic result that extremal black holes cannot be overcharged. For nearly extremal black holes, we also generalize Hubeny's scenario by showing that overcharging is possible in a small region of parameter space. We check how $D$ affects the overcharging parameter space, and find that overcharging becomes increasingly difficult for nearly-extremal black holes in the large-$D$ limit.
\end{abstract}


\maketitle

\section{Introduction}
The cosmic censorship conjecture \cite{penrose1969} is a long-standing open question in classical general relativity that continues to attract interest. Loosely speaking, the conjecture asserts that singularities arising from gravitational collapse are always hidden behind event horizons. It is useful in so much as it affords general relativity predictability as a classical theory. Its violation would compel input from a more complete and as-yet-undiscovered theory -- quantum gravity -- in order to make predictive statements on observables pertaining to events at or near singularities. 
 
That it remains a challenging open conjecture is due in part to its lack of a rigorous formulation. In spite of this, or perhaps because of this, there has been no shortage of efforts to find counterexamples  \cite{semiz2011dyonic,toth2012test,duztacs2014electromagnetic,roberts1989scalar,natario2016test,zhang2014testing,campos2016cosmic,richartz2011challenging,gao2008collapsing,wald1974gedanken,needham1980cosmic,saa2011destroying,gao2013destroying,chirco2010gedanken,hubeny1999overcharging,jacobson2009overspinning,li2013destroying,gwak2016cosmic,zimmerman2013self,colleoni2015self}. Nonetheless, the cosmic censorship conjecture is widely believed to be true, and the challenge to those who adhere to it is to explain why the counterexamples flounder upon closer scrutiny, and more importantly, to understand the mechanism that enforces cosmic censorship. Most of these counterexamples require some degree of fine-tuned initial data. An exception, however, can be found in higher dimensions. Higher-dimensional ($D > 4$) black hole analogs such as black strings, black rings, and \textit{p}-branes have long been demonstrated to be unstable to small perturbations, and these have recently been shown to lead to the formation of a naked singularity \cite{gregory1993black,lehner2010black,Figueras}. The end state of this instability is the only known generic violation of cosmic censorship so far. 

A common approach to unveiling the singularities of black holes is by making them absorb point particles with certain properties. The seminal work by Wald \cite{wald1974gedanken} was the first to seriously explore this possibility. Black hole overcharging is the process by which a test particle of sufficient mass $m$, charge $q$, and energy $E$ falls down a charged black hole of mass $M$ and charge $Q$, overcomes the electrostatic repulsion between them, and is absorbed by the black hole, ultimately resulting in a spacetime that no longer corresponds to a black hole but a naked singularity. Wald demonstrated that when one tries to overcharge an extremal Kerr-Newman black hole in this way, however, the electrostatic repulsion prevents the particle from crossing the event horizon \cite{wald1974gedanken}. Many years later, Hubeny \cite{hubeny1999overcharging} revisited this scenario and discovered that if one starts with a nearly-extremal Reissner-Nordstr\"{o}m black hole instead of an extremal one, then overcharging is possible.

The same situation holds for spinning black holes. A black hole is overspun when it absorbs a test body with enough angular momentum such that the resulting metric after absorption is that of a naked singularity. Again, in \cite{wald1974gedanken}, it was demonstrated that overspinning is impossible for an extremal Kerr-Newman black hole because the test body cannot overcome the spin-spin repulsion between the particle and the black hole. In the same spirit as Hubeny, Jacobson and Sotiriou \cite{jacobson2009overspinning} found that starting with near-extremality evades this restriction. 

Overcharging and overspinning share the same fine-tuning flaw of other mechanisms for creating naked singularities; only for an infinitesimally narrow region of parameter space do they succeed. A more serious shortcoming though, already acknowledged by \cite{hubeny1999overcharging} and \cite{jacobson2009overspinning}, is that both analyses rely on the test-particle approximation. It has been widely believed then that finite-charge and finite-mass corrections to the dynamics of the point particle drastically affect the outcomes of both scenarios \footnote{In \cite{hubeny1999overcharging}, the impact of the local electromagnetic self-force was assessed and shown not to prevent overcharging. However, this was a very tentative result because the local approximation to the self-force is known to be quite inaccurate.}. These corrections are known as self-force effects. The self-force on a particle moving in a curved spacetime arises from the interaction between a particle and the fields it produces. In general, it cannot be computed straightforwardly \cite{vega2011effective, wardell2012generic,vega2009self,wardell2009,barack2000mode,detweiler2003self}, though the technology for such computations has progressed tremendously in recent years and remains an active area of research \cite{poisson2011motion}. 

A more complete picture of the influence of the  electromagnetic self-force on the Hubeny overcharging scenario was revealed by the work of Zimmerman et. al. \cite{zimmerman2013self}. There it was shown that the self-force prevents the test particle from crossing the event horizon, becoming strongly repulsive as the charged particle gets close to the event horizon, and creating a turning point in the trajectory of the particle just as it is about to overcharge the black hole. Colleoni et. al. reach similar conclusions in their careful study of the inclusion of gravitational self-force effects in the Kerr overspinning scenario \cite{colleoni2015overspinning}. The emerging picture thus confirms the expectation that four-dimensional black holes are immune to overcharging or overspinning by point particles, and that indeed it is the self-force that acts as the cosmic censor in these scenarios. 

In this short paper, we contribute to this developing narrative by extending the Hubeny overcharging scenario to higher-dimensional black holes. Indeed, the fact that generic violations of cosmic censorship occur in the nonlinear evolution of higher-dimensional black objects suggests that higher dimensions might be a more fertile arena for seeking out violations. However, earlier work on extending the overspinning process to higher dimensions by Bouhmadi-Lopez et. al. conclude that potentially destructive point particles with large angular momenta are not captured by the extremal Myers-Perry black holes. Higher-dimensional overspinning in the extremal case cannot succeed. To the best of our knowledge, the analogous overcharging scenario is yet to be extended to dimensions $D > 4$. We find that, just like in four dimensions, overcharging an extremally charged black hole is impossible in the test particle limit.  We also show that Hubeny's conclusions in $D=4$ extend to higher dimensions as well: there exist charged test particles that can overcharge a nearly-extremal charged Schwarzschild-Tangherlini black hole. 

Very recent work by Sorce and Wald \cite{sorce_wald} show quite generally that overcharging and overspinning of nearly-extremal black holes cannot occur, though it is not clear if this statement extends to all dimensions.

The rest of the paper proceeds as follows. To set the stage, we first briefly review the Hubeny overcharging scenario in a nearly-extremal Reissner-Nordstr\"{o}m black hole. We then look at generalizing this situation to a charged Schwarzschild-Tangherlini black hole. We work out the kinematics of charged particle infall in this background. From these we derive the conditions for overcharging these black holes -- what we call generalized Hubeny inequalities -- and show that these cannot be satisfied when the black hole is extremal, but can be satisfied for a small region of parameter space when the black hole is nearly-extremal. These generalize the Wald and Hubeny results to $D>4$. We also check how the overcharging parameter space depends on $D$. Finally, we summarize with a discussion of our results.

Throughout this paper our metric signature is mostly plus $(-,+,+,\ldots, +,+)$. For consistency with past work in $D=4$, we adopt geometric units in which $G_D=c=1$.

\section{The Hubeny scenario}

We first recall the Reissner-Nordstr\"{o}m (RN) line-element in the usual Schwarzschild coordinates, which describes the spacetime of a charged, asymptotically-flat solution to the Einstein-Maxwell equations. The line-element reads
\begin{equation}
ds^2= - f(r) dt^2 +f(r)^{-1} dr^2 + r^2 d\Omega^2
\end{equation}
where
\begin{equation}
f(r) = 1-\dfrac{2M}{r} + \dfrac{Q^2}{r^2},
\end{equation}
and $d\Omega^2= d\theta^2+\sin^2\theta d\phi^2$ is the metric on the unit two-sphere.
This solution represents a black hole of mass $M$ and charge $Q$ possessing an event horizon at $r=r_+:=M+\sqrt{M^2-Q^2}$. The black hole supports an electromagnetic field and four-potential whose only non-zero components are $F_{tr}=-Q/r^2$ and $A_t = -Q/r$, respectively.

When $Q>M$, there is no event horizon, and the Reissner-N\"{o}rdstrom solution bears its curvature singularity located at $r=0$ to the outside universe. The case $Q=M$ represents an extremal RN black hole, and when  $(Q-M)/M \ll 1$, we have a nearly-extremal RN black hole. Hubeny parametrizes near-extremality by relating the mass and charge of the black hole as $Q=M-2\epsilon^2$ and requiring that $0< \epsilon \ll 1$. Extremal RN black holes are then those for which $\epsilon =0$. 

The Hubeny scenario consists of a test charge, with mass $m$ and charge $q$, falling radially towards a nearly-extremal Reissner-Nordstr\"{o}m black hole. This radial infall proceeds according to the equation of motion
\begin{equation}
ma^\alpha = q F^{\alpha}{}_{\beta}u^\beta,
\end{equation}
whereby the charged particle is met by the electrostatic repulsion coming from the black hole. For overcharging to occur, the charged particle must overcome this repulsion and accrete its energy $E$ and charge $q$ onto the black hole so that the latter's final mass and charge are $M+E$ and $Q+q$, respectively. When 
\begin{equation}
\label{eq:overcharge}
Q+q > M+E,
\end{equation}
the final state is said to be overcharged. 

For the particle to cross the event horizon, its four-velocity, $u^\alpha = \left(\dot{t},\dot{r},0,0\right)$, must satisfy the conditions
\begin{itemize}
\item[(i)] $\dot{r}^2>0 \hspace{0.2cm} \forall r \geq r_+$, and
\item[(ii)] $\dot{t}>0 \hspace{0.2cm} \forall r>r_+$.
\end{itemize} 
The overdot means differentiation with respect to the particle's proper time.  Condition (i) simply means that no turning point exists in its trajectory  before it enters the horizon. The particle must have sufficient energy $E$ to overcome electrostatic repulsion.  Condition (ii) ensures that the four-velocity is future-pointing at the horizon. Taking Eq.~(\ref{eq:overcharge}) and conditions (i) and (ii) together, one can derive the following inequalities
\begin{subequations}
\label{eq:ineq}
	\begin{align}
   	q>\frac{r_+-Q}{2} \label{q_Hubeny},  \hspace{1cm}\\
   	\dfrac{qQ}{r_+}<E<q+Q-M \label{E_Hubeny}, \\
    m<Q\sqrt{\dfrac{2MEq-Q(E^2+q^2)}{Q(M^2-Q^2)}} \label{m_Hubeny}.
	\end{align}
\end{subequations}
These are the conditions for overcharging, and are constraints on the particle parameter space $\lbrace m, E, q\rbrace$. Particles with mass, energy, and charge $\lbrace m, E, q\rbrace$ within this domain are then said to produce an overcharged final state, i.e. a naked singularity, from a RN black hole of mass and charge $\{M,Q\}$.

In addition to these conditions, one must bear in mind that the test particle approximation has to be valid, i.e., the test particle's stress-energy tensor must not significantly disturb the background spacetime. Thus, the particle parameters  $\{m, E,q\}$ have to be much smaller than black hole parameters $\{M,Q\}$.

In the extremal limit, the inequalities (\ref{q_Hubeny})-(\ref{m_Hubeny}) reduce to 
\begin{subequations}
\begin{align}
   	q &>0,\\
   	q < \,&E < q \\
    m &< \infty,
\end{align}
\end{subequations}
which admit no solution. We therefore recover Wald's result that extremal RN black holes cannot be overcharged.

If the RN black hole is nearly-extremal, so that $Q= M-2\epsilon^2$, Hubeny showed how to obtain a solution to Eqs.~(\ref{q_Hubeny})-(\ref{m_Hubeny}). In particular, for any nonzero $\epsilon \ll 1$, the choices
\begin{subequations}
\begin{align}
q & = a\epsilon,\,\,\,\, a>1, \\ 
E &= a\epsilon - 2b\epsilon^2, \,\,\,\, 1<b<a, \\
m &= c\epsilon, \,\,\,\, c< \sqrt{a^2-b^2}
\end{align}
\end{subequations}
will satisfy the above inequalities \cite{hubeny1999overcharging,poissonnotes}. 

\section{Overcharging in $D$ dimensions}

In this section, we show that in $D$ dimensions the Hubeny inequalities generalize to 
\begin{equation}
q>r_+^{D-3}\left(\dfrac{M-\omega_D Q}{\omega_D r_+^{D-3}-Q/(D-3)}\right),
\end{equation}
\begin{equation}
\dfrac{qQ}{(D-3)r_+^{D-3}}<E<\omega_D\left( Q+q \right)-M,
\end{equation}
and,
\begin{equation}
m<\omega_D Q \sqrt{\dfrac{2MEq-Q(E^2+\omega_D^2q^2)}{Q(M^2-\omega_D^2 Q^2)}},
\end{equation}
where
\begin{equation}
r_+^{D-3}  = \dfrac{M}{(D-3)\omega_{D}^2}\left(1+ \sqrt{1-\dfrac{\omega_D^2Q^2}{M^2}} \right),
\end{equation} and $\omega_D$ is defined as 
\begin{equation}
\omega_D = \sqrt{\dfrac{(D-2)}{(D-3)}\dfrac{\Omega_{(D-2)}}{8\pi }}.
\end{equation}
It is easy to check that $\omega_D=1$ in $D=4$ and that these inequalities correctly reduce to the Hubeny inequalities in $D=4$.

\subsection{Einstein-Maxwell action}
The Einstein-Maxwell equations in $D$ dimensions arise from the action
\begin{equation}
S = S_{\textrm{g}} + S_{\textrm{EM}} + S_{\textrm{m}} + S_{\textrm{int}}
\end{equation}
where
\begin{align}
S_{\textrm{g}} &= \dfrac{1}{4 \Omega_{(D-2)}G_D} \int d^Dx \sqrt{-g}\,\, R, \\
S_{\textrm{EM}} &= -\dfrac{1}{4 \Omega_{(D-2)}} \int d^Dx \sqrt{-g}\,\, F^{\mu\nu}F_{\mu\nu}, \\ 
S_{\textrm{m}} &= -m\int_\gamma d\tau, \\
S_{\textrm{int}} &= \int d^Dx \sqrt{-g}\,\, A_\mu j^\mu.
\end{align}
For a particle with charge $q$ moving along a worldline $\gamma$, which is parametrized by $x^\mu = z^\mu(\lambda)$ for some arbitrary parameter $\lambda$, 
\begin{equation}
j^\mu = q \int_\gamma d\lambda\,\,\dfrac{dz^\mu}{d\lambda} \,\,\dfrac{\delta^{(D)}(x^\mu-z^\mu(\lambda))}{\sqrt{-g}}.
\end{equation}
This action yields the field equations
\begin{equation}
G_{\alpha\beta} = 8\pi G_D T_{\alpha\beta},
\end{equation}
\begin{equation}
F^{\alpha\beta}_{;\beta} = \Omega_{(D-2)}j^\alpha,
\end{equation}
\begin{equation}
ma^\alpha = qF^\alpha{}_\beta u^\beta.
\end{equation}
upon imposing stationarity of the action with respect to $g_{\alpha\beta}$, $A_\mu$ and $z^\mu(\lambda)$, respectively.The stress-energy tensor is given by 
\begin{equation}
T_{\alpha\beta} = \dfrac{1}{\Omega_{(D-2)}}\left(F_{\alpha\mu}F^\mu_\beta- \dfrac{1}{4} g_{\alpha \beta}F^{\mu\nu}F_{\mu\nu}\right).
\end{equation}

\subsection{Charged Schwarzschild-Tangherlini black holes in $D$ dimensions}

The  $D$-dimensional analogue to the Reissner-Nordstr\"{o}m solution, or the charged Schwarzschild-Tangherlini solution, is again parametrized by a mass $M$ and charge $Q$. Its line element is given by
\begin{equation}
ds^2 = -f(r) dt^2 +f(r)^{-1}dr^2 +r^2 d\Omega_{D-2}^2
\end{equation}
where
\begin{equation} 
f(r) = 1-\frac{\mu}{r^{D-3}}+\frac{\xi^2}{r^{2(D-3)}},
\end{equation}
and
\begin{align}
\mu &= \dfrac{16\pi M}{(D-2)\Omega_{(D-2)}},\\
\xi &= \left(\dfrac{8\pi}{\Omega_{(D-2)}(D-2)(D-3)}\right)^{1/2}Q,
\end{align}
with $M$ and $Q$ being the ADM mass and charge of the black hole.
Finally,
\begin{equation}\nonumber
d\Omega_{D-2}^2 = d\theta_{1}^2 + \sin^2\theta_{1}d\theta_{2}^2  + \dotsc +  \sin^2\theta_{1}\dotsb \sin^2\theta_{D-3}d\theta_{D-2}^2 
\end{equation}
is the line-element of the unit $(D-2)$-sphere of volume $\Omega_{(D-2)} = 2 \pi^{(D-1)/2}/\Gamma\left( (D-1)/2\right)$. When $D=4$, this solution reduces to the expected Reissner-Nordstr\"{o}m solution. This solution supports an electromagnetic field and potential whose only non-zero components are $F_{tr}=Q/r^{D-2}$ and $A_t = -Q/((D-3)r^{D-3})$.

The metric function $f$ has an outer root 
\begin{equation}
r_+^{D-3} = \dfrac{\mu}{2}\left(1+\sqrt{1-\dfrac{4\xi^2}{\mu^2}}\right),
\end{equation}
which locates the event horizon. The location of the event horizon in terms of the black hole mass and charge is
\begin{equation}
r_+^{D-3}  = \dfrac{8\pi M}{(D-2)\Omega_{(D-2)}}\left(1+ \sqrt{1-\dfrac{(D-2)}{(D-3)}\dfrac{\Omega_{(D-2)}}{8\pi }\dfrac{Q^2}{M^2}} \right).
\end{equation}
While taking note of the definition of $\omega_D$ as
\begin{equation}
\omega_D = \sqrt{\dfrac{(D-2)}{(D-3)}\dfrac{\Omega_{(D-2)}}{8\pi }},
\end{equation}
we can rewrite the location of the horizon as
\begin{equation}
r_+^{D-3}  = \dfrac{M}{(D-3)\omega_{D}^2}\left(1+ \sqrt{1-\dfrac{\omega_D^2Q^2}{M^2}} \right).
\end{equation} 
An event horizon exists only when
\begin{equation}
\label{eq:ehd}
M \geq \omega_D Q.
\end{equation}
The extremal state occurs when the equality holds in relation ~(\ref{eq:ehd}). The overcharged state occurs when this is violated or when
\begin{equation}
Q > \omega_D^{-1} M.
\end{equation}

\subsection{Crossing conditions for charged particle infall}
The set-up for overcharging proceeds exactly as in Hubeny. We consider a particle of a certain mass $m$ and charge $q$ radially falling into the black hole with just the right parameters so that it crosses the horizon $r_+$ and produces a spacetime that violates Eq.~(\ref{eq:ehd}). Our goal now is to generalize the Hubeny inequalities in Eqs.~(\ref{q_Hubeny})-(\ref{m_Hubeny}) to $D$ spacetime dimensions.

The equation of motion for the charged particle remains $ma^\alpha = q F^\alpha{}_\beta u^\beta$. For a radial trajectory $z^\alpha=(T(\tau),R(\tau),0,\dotsc,0)$, where \(\tau\) is the proper time, the point charge has a velocity
	\begin{equation}\label{fourv}
	u^{\alpha}=\frac{dz^\alpha}{d\tau}=(\dot{T}(\tau),\dot{R}(\tau),0,\dotsc,0),
	\end{equation}
and momentum given by
	\begin{equation}
	p_\alpha=\left(-mf\dot{T}-\frac{qQ}{(D-3)r^{D-3}},f^{-1}\dot{R},0,\dotsc,0\right).	\end{equation}
Associated with the time-like Killing vector of the spacetime, $\xi_{(t)}^\alpha=(1,0,\dotsc,0),$ is a constant of motion given by 	\begin{equation}\label{Energyequation}
	E=-p_\alpha\xi^\alpha_{(t)}=mf\dot{T}+\frac{qQ}{(D-3)r^{D-3}}.
	\end{equation}
From this, we get
	\begin{equation}
    \label{tdoteqn}
	\dot{T}=\frac{1}{mf}\left(E-\frac{qQ}{(D-3)r^{D-3}}\right). 
	\end{equation} 
Moreover, from the normalization of the velocity $u_\alpha u^\alpha =-1$, the equation of motion for $R(\tau)$ becomes
	\begin{equation}\label{rdot2eqn}
	\dot{R}^2=\frac{1}{m^2}\left(E-\frac{qQ}{(D-3)r^{D-3}}\right)^2-f(r).
	\end{equation}
For the particle to cross the horizon, it is sufficient to require that $\dot{T} > 0$ for $r>r_+$ and $\dot{R}^2>0$ for all $r \geq r_+$. Evaluating Eq.~(\ref{rdot2eqn}) at $r=r_+$, we get 
\begin{equation}\label{Elower}
E> \dfrac{qQ}{(D-3)r_+^{D-3}}.
\end{equation}
From Eq.~(\ref{rdot2eqn}), the condition $\dot{R}^2 > 0$ for all $r>r_+$ can be written as
\begin{equation}
\label{mboundnew}
m^2<\dfrac{1}{f(r)}\left(E-\frac{qQ}{(D-3)r^{D-3}}\right)^2, \forall \,\,r>r_+.
\end{equation}
The minimum of the right hand side  occurs at 
\begin{equation}\label{mboundmin}
	r_m^{D-3}=\dfrac{mQq-Q^2E}{(D-3)\omega_D^2  qQ - (D-3) ME}.
	\end{equation}
Substituting this back into Eq.~{\ref{mboundnew}}, we get
\begin{equation}\label{mbound_final}
m<\omega_D Q \sqrt{\dfrac{2MEq-Q(E^2+\omega_D^2q^2)}{Q(M^2-\omega_D^2 Q^2)}}.	\end{equation}
To summarize, the crossing conditions $\dot{T} > 0$ and $\dot{R}^2 > 0$ for all $r>r_+$ lead to the two inequalities
\begin{subequations}
\begin{align}
E &> E_{\textrm{min}} := \dfrac{qQ}{(D-3)r_+^{D-3}},\label{eq:cross1}\\
m &< m_{\textrm{max}} := \omega_D Q \sqrt{\dfrac{2MEq-Q(E^2+\omega_D^2q^2)}{Q(M^2-\omega_D^2 Q^2)}}. \label{eq:cross2}
\end{align}
\end{subequations}
The inequalities above guarantee that a charged particle characterized by $\{m,E,q\}$ will cross the event horizon. Now we seek to ascertain the form of the $D$-dimensional RN metric after it absorbs the charged particle.

\subsection{Overcharging condition}
Like all previous work \cite{wald1974gedanken,hubeny1999overcharging,jacobson2009overspinning,bouhmadi2010black}, we assume that the particle energy $E$ fully accretes to the ADM mass of the black hole. The ADM mass upon absorption of the particle then simply increases as $M \rightarrow M+E$ while the ADM charge increases as $Q$ to $Q+q$. This assumption misses out on all radiative/self-force effects, which of course lie outside the test particle approximation.

The line element of the spacetime after absorption of the charged particle becomes
	\begin{subequations}\label{metricwithparticle}
    \begin{equation}
ds^2 = -f(r) dt^2 +f(r)^{-1}dr^2 +r^2 d\Omega_{D-2}^2, 
\end{equation}
where
\begin{equation}
\begin{split}
f(r)=1 &-\frac{16\pi}{(D-2)\Omega_{D-2}}\frac{M+E}{r^{D-3}} \\
&+ \frac{8 \pi}{\Omega_{(D-2)}(D-2)(D-3)}\frac{(Q+q)^2}{r^{2(D-3)}}.
\end{split}
\end{equation} 
\end{subequations}
The location of the horizons for this new line element \eqref{metricwithparticle} is given by
\begin{equation}
r_{\pm}^{D-3}  = \dfrac{M}{(D-3)\omega_D^2}\left(1 \pm \sqrt{1-\dfrac{\omega_D^2(Q+q)^2}{(M+E)^2}} \right).
\end{equation}

The line element describes the spacetime around a black hole when $r_{\pm}$ are real or 
	\begin{equation}\label{BHcondition}
	M+E \geq \omega_{D}\left( Q+q \right).
\end{equation}
A naked singularity is described by the RN line element when 
	\begin{equation}\label{OC}
	Q+q > \omega_D^{-1}\left( M+E \right).
	\end{equation} 
This can be rewritten to give an upper bound for $E$
\begin{equation}\label{Eupper}
	E < E_{\textrm{max}} :=\omega_D \left( Q+q \right)-M.
\end{equation}
Together with Eq.~(\ref{Elower}), Eq.~(\ref{Eupper}) can be written as
\begin{equation}\label{Econstraint}
\dfrac{qQ}{(D-3)r_+^{D-3}}<E<\omega_D\left( Q+q \right)-M,
\end{equation}
allowing us to derive an upper bound for the charge,
\begin{equation}
\dfrac{qQ}{(D-3)r_+^{D-3}}<\omega_D \left( Q+q \right)-M, 
\end{equation}
or
\begin{equation}\label{qlower}
q> q_{\textrm{min}}:= r_+^{D-3}\left(\dfrac{M-\omega_D Q}{\omega_D r_+^{D-3}-Q/(D-3)}\right).
\end{equation}
This completes the derivation of the generalized Hubeny inequalities. 

\subsection{Overcharging extremal black holes}
We now use these inequalities to constrain the parameter space for a charged particle that is about to fall towards an extremal black hole. In the extremal limit, $M = \omega_D Q$, the horizon location becomes
\begin{equation}
r^{D-3} = r_+^{D-3} = \dfrac{M}{(D-3)\omega_D^2} = \dfrac{Q}{(D-3) \omega_D},
\end{equation}
and the system of inequalities reduces to
\begin{align}
q &>0, \\
q<& E<q, \\
m &<\infty,
\end{align} which does not have a solution. This confirms that Wald's result remains correct for higher-dimensional extremal black holes.  

\subsection{Overcharging in the nearly-extremal case}

Looking now at the near-extremal case, we note that near-extremality in $D$-dimensions can be parameterized using an extremality parameter $0<\epsilon \ll 1$ as
\begin{align}
M \equiv 1, \,\,\,\, Q \equiv \omega_D^{-1}-2\epsilon^2.
\label{nearextremecond}
\end{align}
The expression for the event horizon then reduces to
\begin{equation}
r_{+}^{D-3} = \dfrac{1+2\sqrt{\omega_D}\epsilon}{(D-3)\omega_D^2},
\end{equation}
while the lower bound for $q$ becomes 
\begin{subequations}
\begin{align}
q>\dfrac{\epsilon + 2\sqrt{\omega_D}\epsilon^2}{\sqrt{\omega_D}+\omega_D\epsilon}.
\end{align}
\end{subequations}
Expanding in $\epsilon$, this becomes
\begin{equation}
q> \omega_D^{-1/2} \epsilon + \epsilon^2 +\mathcal{O}\left( \epsilon^3 \right),
\end{equation}
which can be satisfied by the choice
\begin{equation} \label{qnearextreme}
q = A \epsilon, \,\,\, A > \omega_D^{-1/2}.
\end{equation}
At $D=4$, this reduces to
\begin{equation}
q=a\epsilon, \,\,\,\, a>1,
\end{equation}
the same solution to the Hubeny inequalities in the near-extremal case in the lowest order. 

\begin{figure}
\includegraphics[scale=0.6]{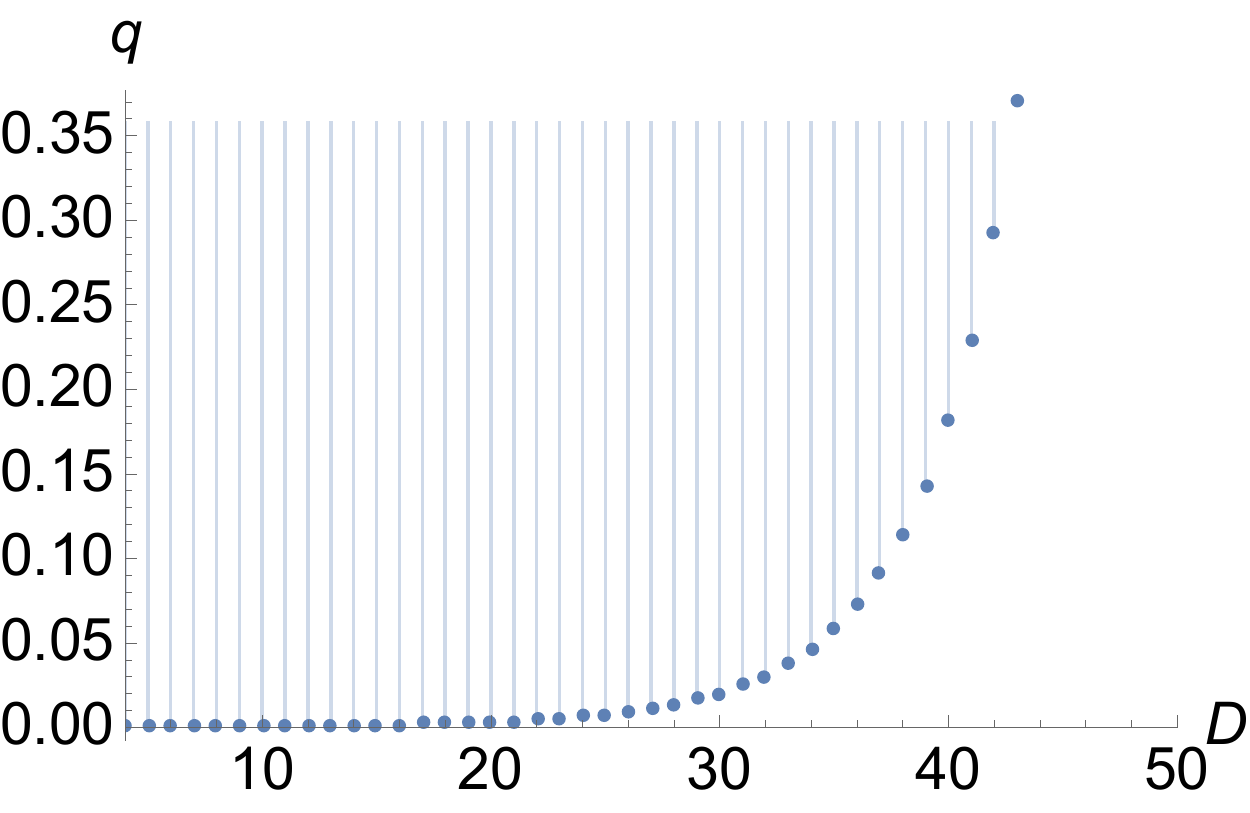}
\centering
\caption{Allowed parameter space in $q$ for nearly-extremal BH with $\epsilon=0.001$. The dots represent  $q_{\textrm{min}}$, the smallest allowed charge for overcharging to occur.  As $D \rightarrow \infty$, $q_{\textrm{min}} \rightarrow \infty$.} 
\label{qmin_plot}
\end{figure}

Inserting Eq.~(\ref{qnearextreme}) to the energy constraints, Eqs.~(\ref{Elower}) and ~(\ref{Eupper}), we get
\begin{subequations}
\begin{align}
E<\omega_D \left(A  \epsilon - 2\epsilon^2 \right),
\end{align}
\begin{align}
E>\omega_D A  \left(  \epsilon - 2\epsilon^2 \sqrt{\omega_D}+2\epsilon^3  + \mathcal{O}\left(\epsilon^4\right) \right),
\end{align}
\end{subequations}
which can be satisfied with the choice
\begin{equation}\label{Enearextreme}
E = \omega_D \left(A  \epsilon - 2 B \epsilon^2\right), \,\,\, 1<B<\sqrt{\omega_D}A.
\end{equation}
This also reduces to the $D=4$ case where
\begin{equation}
E = a \epsilon -2 b\epsilon^2, \,\,\,1<b<a.
\end{equation}
Inserting Eqs.~(\ref{nearextremecond}),(\ref{qnearextreme}), and (\ref{Enearextreme}) into Eq.~\eqref{mbound_final}, we get 
\begin{equation}
m< \epsilon\sqrt{A^2\omega_D^2-B^2\omega_D}  - \epsilon^2\dfrac{AB \omega_D^{3/2}}{\sqrt{A^2 \omega_D - B^2}} + \mathcal{O}(\epsilon^3).
\end{equation}
This can be satisfied with the choice 
\begin{equation}
m = C\epsilon, \, \,\,\, C<\sqrt{A^2\omega_D^2 -B^2\omega_D}.
\end{equation}
Again, this reduces to Hubeny's mass constraint in $D=4$. 
\begin{equation}
m = c\epsilon, \,\,\,\, c<\sqrt{a^2-b^2}. 
\end{equation}
This demonstrates that it is quite easy to find a solution to the generalized Hubeny inequalities for the case of a nearly-extremal black hole.

\begin{figure}

\includegraphics[scale=0.6]{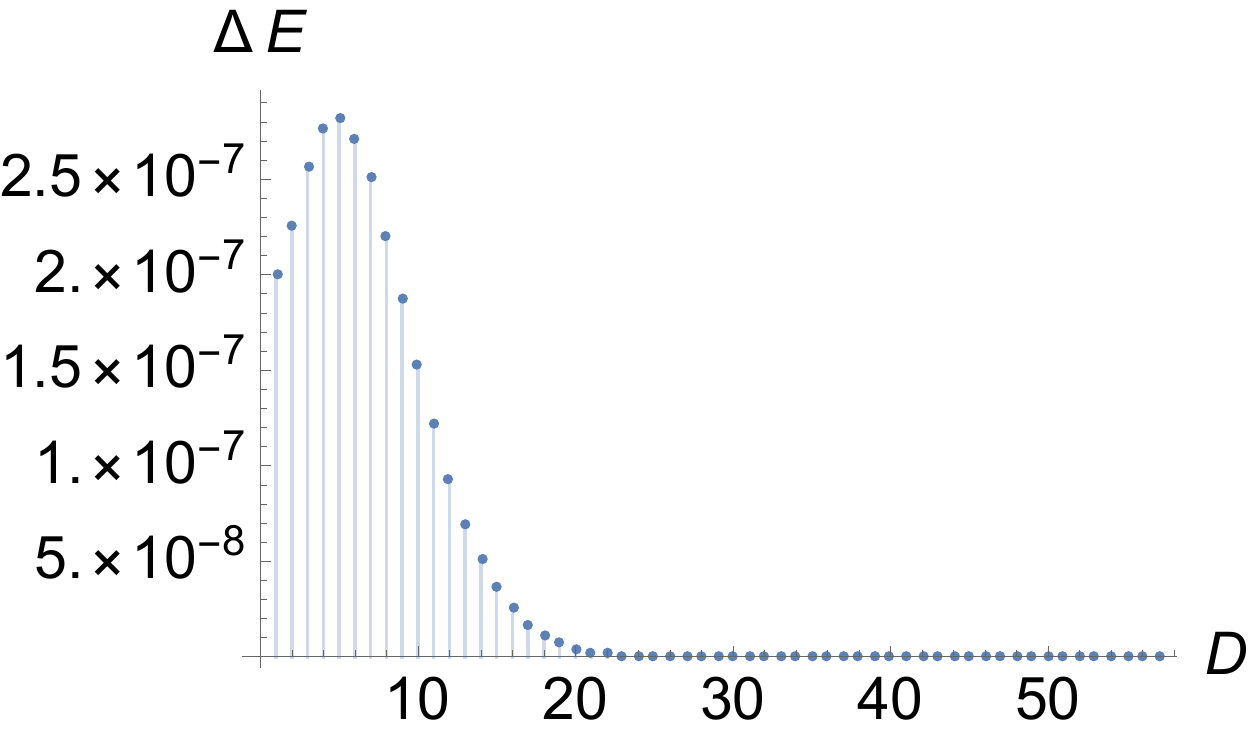}
\caption{Width of parameter space in $E$ for nearly-extremal BH with $\epsilon=0.001$. The dots represent the value of $\Delta E$. As $D \rightarrow \infty$, $\Delta E \rightarrow 0$.} 
\label{E_diffplot}
\centering
\end{figure}

\begin{figure}[h]
\includegraphics[scale=0.6]{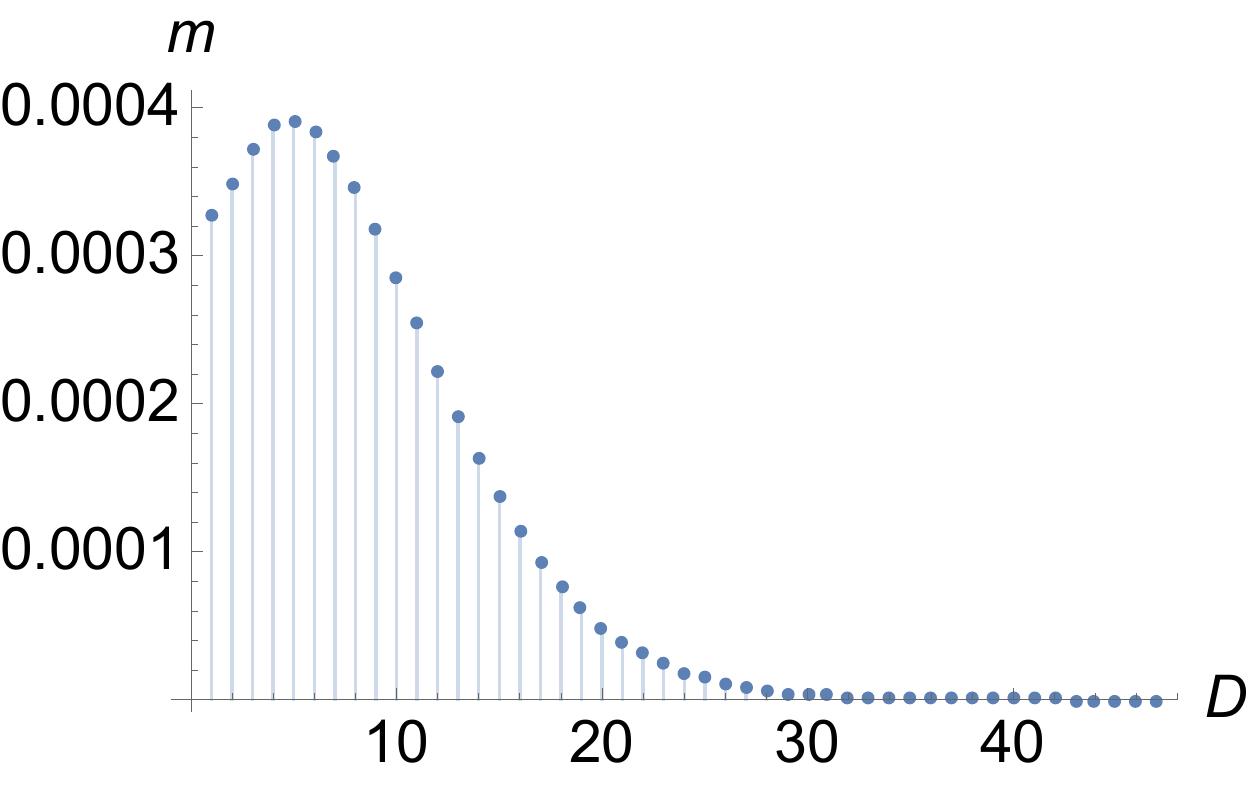}
\caption{Allowed parameter space in $m$ for nearly-extremal BH with $\epsilon=0.001$. The dots represent the value of $m_{\textrm{max}}$, the maximum allowed particle mass for overcharging to occur. As $D \rightarrow \infty$, $m_{\textrm{max}} \rightarrow 0$.}
\label{mplot}
\centering
\end{figure}

Looking now at what happens when $D \rightarrow \infty$, we first notice in Fig.~\ref{qmin_plot} that the smallest charge allowing for overcharging, $q_{\textrm{min}}$, increases with $D$. Therefore, the required charge eventually becomes too large to satisfy the test particle assumption, $q\ll M\sim Q$. In the energy sector, Fig.~\ref{E_diffplot}, the allowed energy range, $\Delta E=E_{\textrm{max}}-E_{\textrm{min}}$, is found to shrink to zero as $D\rightarrow \infty$. In Fig.~\ref{mplot}, the maximum allowed mass is seen to go to zero in the same limit. We conclude from this that going to higher dimensions makes overcharging 
systematically more difficult. 

\section{Summary and Conclusion}

The goal in this work was to explore the possibility of overcharging of black holes by point particles in higher dimensions. To this end, we studied the radial infall of a charged particle in a charged, spherically-symmetric, $D$-dimensional black hole. As in Hubeny, we reduced the conditions leading to cosmic censorship violation in this scenario into a set of generalized Hubeny inequalities. We further learned that these inequalities cannot be satisfied in extremal black holes, but that they can be satisfied when the black holes are nearly-extremal. Thus, the results of Wald and Hubeny for $D=4$ remain true in higher dimensions.

These results are not entirely surprising. The interaction between a charged black hole and an infalling test charge consists of a competition between their gravitational attraction and electromagnetic repulsion. Wald's classic no-go result in the extremal case ($Q=M$) can be taken to mean that these competing effects precisely cancel. On the other hand, starting with a black hole charge just shy of extremality (i.e. $Q=M(1-2\epsilon^2$)) weakens the electromagnetic  repulsion sufficiently enough to allow some test charges to cross the horizon and overcharge the black hole. In higher dimensions, the strengths of both gravitational attraction and electromagnetic repulsion scale precisely in the same way as $\sim r^{-(D-2)}$. So the previous considerations concerning the balance between these two effects can be expected to remain true. 

There may be little doubt that our conclusions are an artifact of the test particle approximation, just as in $D=4$.   Destroying the event horizon is a violation not only of cosmic censorship but also the second and third laws of black hole mechanics, which ought to hold in higher dimensions \cite{emparan2008black}. Beyond these theoretical considerations, for any process that violates cosmic censorship, we are always left to ask about what possible mechanism might prevent the violation. For scenarios involving point particles, the natural choice of cosmic censor is the back-reacting self-force, the calculation of which, in higher dimensions, is a subject still in its infancy, and is an active area of research \cite{taylorflanagan,beach2014self,Frolov2014,harte2016}. Our work can be viewed as a strong invitation to pursue higher-dimensional self-force calculations. 

There are indications that the self-force in higher dimensions gets divergently repulsive as the particle approaches the horizon \cite{taylorflanagan}, which would mean that none of the charged particles we identify as overcharging do, in fact, cross the horizon. However, the matter is far from settled \cite{beach2014self}. It would be interesting to see how this story unfolds, as self-force calculations in higher dimensions mature to the state that has been reached in $D=4$. We leave this problem for future work.

\section{Acknowledgements}
The authors thank Eric Poisson for helpful comments that helped shape this work. This research is supported by the University of the Philippines OVPAA through Grant No.~OVPAA-BPhD-2016-13. KSR acknowledges the financial support of the Philippines' Department of Science and Technology through the Advanced Science and Technology Human Resources Development Program.

\bibliography{overcharginghigherdbib}

\end{document}